\newcommand{\rem}[1]{}
\documentclass[prl,twocolumn,showpacs,preprintnumbers,amsmath,amssymb,floatfix]{revtex4}
\usepackage{hyperref}
\usepackage[dvips]{graphicx}

\begin{document}

\title{
AFM Dissipation Topography of Soliton Superstructures in Adsorbed Overlayers
}

\author{Carlotta Negri$^1$, Nicola Manini$^{1,2}$,
        Andrea Vanossi$^{2,3}$, Giuseppe E. Santoro$^{2,4}$, and Erio Tosatti$^{2,4}$}
\affiliation{
$^1$Dipartimento di Fisica, Universit\`a degli Studi di
  Milano, Via Celoria 16, 20133 Milano, Italy \\
$^2$International School for Advanced Studies (SISSA)
and CNR-INFM Democritos National Simulation Center, Via Beirut 2-4, I-34151
Trieste, Italy \\
$^3$CNR-INFM National Research Center S3 and Department of Physics, \\
University of Modena and Reggio Emilia, Via Campi 213/A, 41100 Modena,
Italy \\
$^4$International Centre for Theoretical Physics (ICTP),
Strada costiera 11, I-34151 Trieste, Italy
}

\begin{abstract}
In the atomic force microscope, the nanoscale force topography of
even complex surface superstructures is extracted by the changing vibration
frequency of a scanning tip.
An alternative dissipation topography with similar or even better contrast
has been demonstrated recently by mapping the $(x,y)$-dependent tip 
{\em damping}: but the detailed damping mechanism is still unknown.
Here we identify two different tip dissipation mechanisms: local mechanical
softness, and hysteresis.  Motivated by recent data, we describe both of
them in a one-dimensional model of Moir\'e superstructures of
incommensurate overlayers.
Local softness at ``soliton'' defects yields a dissipation contrast that
can be much larger than the corresponding density or corrugation
contrast. At realistically low vibration frequencies, 
however, a much stronger and more effective
dissipation is caused by the tip-induced nonlinear jumping of the soliton,
naturally developing bi-stability and hysteresis. Signatures of this mechanism
are proposed for experimental identification.
%
\end{abstract}

\date{January 6, 2010}


\maketitle

\section{Introduction}
\begin{figure}
\centerline{
{\small (a)}\,\includegraphics[height=35mm,angle=0,clip=]{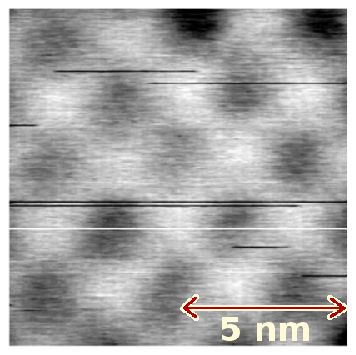}
\hfill
{\small (b)}\,\includegraphics[height=35mm,angle=0,clip=]{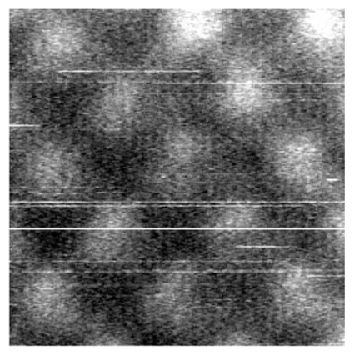}
}
\vskip 2 mm
\centerline{
{\small (c)}\,\includegraphics[height=42mm,angle=0,clip=]{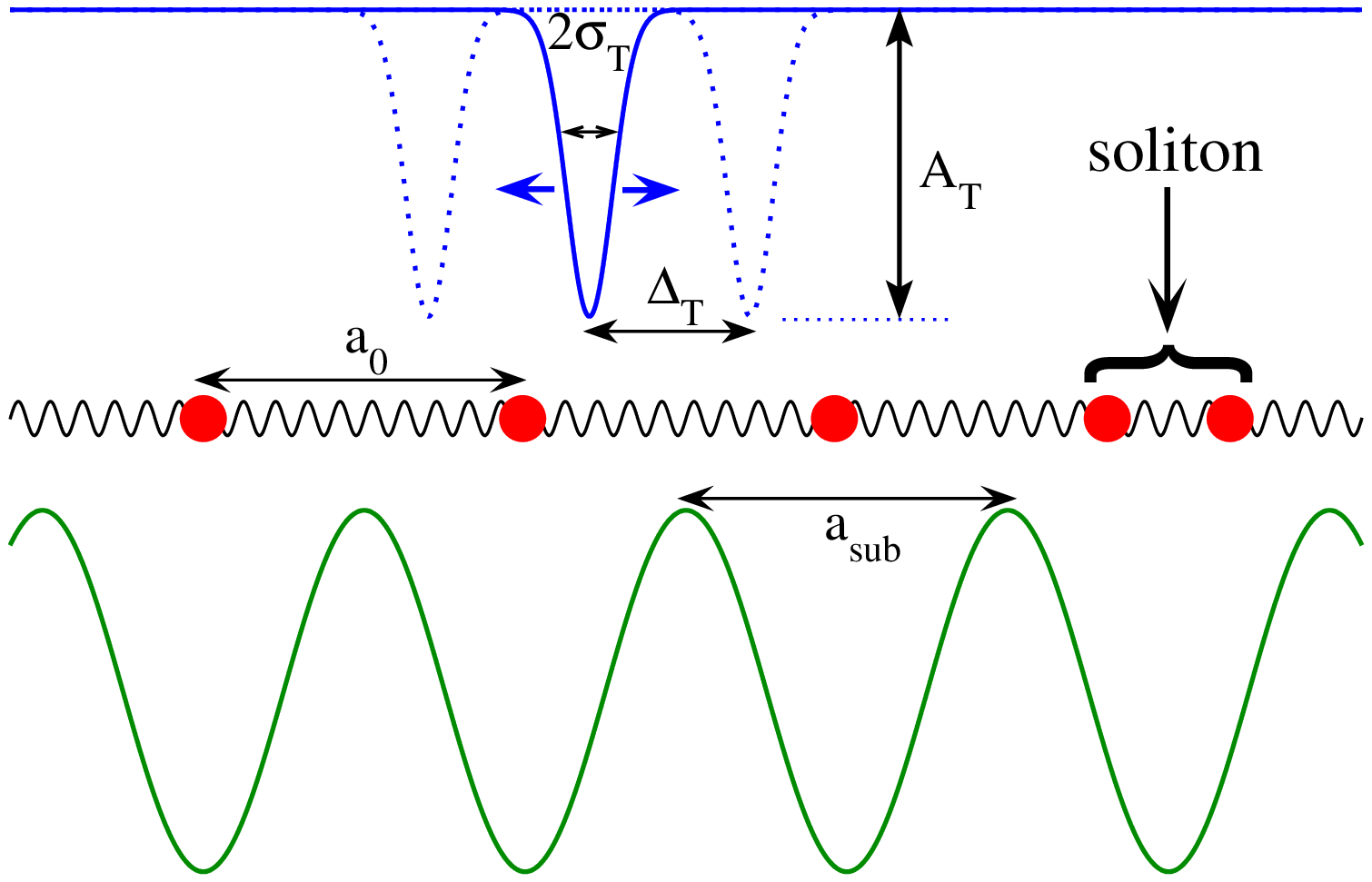}
}
\caption{\label{model:fig} (Color online)
AFM Moir\'e superstructures of incommensurate KBr bilayers
islands/NaCl(100), adapted from Ref.~\cite{Maier07}:
(a) topography;
(b) dissipation. Note the opposite phase: dissipation is largest at the 
soliton, where topographic height is minimal \cite{Baker96}.
(c) 1D simulation model with a rigid substrate potential of period
$a_{\rm sub}$, a harmonic chain of rest length $a_{0}$ (the adsorbate
overlayer), and the localized tip potential $u_{\rm T}$ (here attractive)
oscillating horizontally (in experiments the oscillation is usually
vertical).
}
\end{figure}

%
The tip-based scanning force microscopes of the 
atomic force microscope (AFM) family constitute
perhaps the single most important tool bag in nanotechnology.
The substrate topography  is extracted from a map of the oscillation
frequency of a tip, hovering a short distance above the surface.
Besides the frequency shift however, the tip also develops a damping,
reflecting a position dependent mechanical dissipation.
Maier {\it et al.}\ \cite{Maier08,Maier07} showed recently that AFM
dissipation -- whose general occurrence has been widely discussed by several
groups a decade ago \cite{Gauthier00,Bennewitz00,Loppacher00,LoppacherBis00,Hoffmann01} 
but whose potential importance was still underestimated 
-- is able to map exquisitely delicate
features such as the Moir\'e superstructure pattern formed by misfit
dislocations (``solitons'') of incommensurate KBr adsorbate islands
(Fig.~\ref{model:fig}a) on NaCl(100).
Surprisingly, the experimental dissipation map, Fig.~\ref{model:fig}b, showed similar 
or better contrast than the corresponding topographic map, with a 
characteristic reversed contrast
(higher dissipation at the soliton, where topographic height is minimal
\cite{Baker96}).
Given also the great importance of achieving newer routes toward high-quality
imaging, this is more than a mere curiosity, and deserves a proper
understanding.
Existing linear-response theory and other approaches to AFM dissipation
\cite{Gauthier99,Gauthier00} and to general frictional dissipation
\cite{Granato99,PerssonBook,PerssonTosatti99} suggest a larger tip damping
above softer substrates, and that provides an initial and valuable clue. 
Local tip dissipation can effectively reveal the underlying superstructure, since the
local mechanical compliance is higher for example at surface soliton lines,
where atoms sit at metastable positions.
However at the relatively low AFM oscillation frequencies,
the current understanding 
rules out 
linear response as the chief dissipation mechanism. A typical energy dissipation as
large as $0.01-1$~eV per oscillation can only be accounted for by a
hysteretic response of the interacting tip-substrate system, as was
understood by theoretical analysis \cite{Kantorovich04,Trevethan06,Ghasemi08}, 
and demonstrated experimentally \cite{Schirmeisen05,Hoffmann07}.
Such nonlinear effects of hysteresis 
are most likely involved in the surprisingly large AFM dissipation contrast 
of adsorbate superstructures too. Yet, it is unclear how inert systems (such as for example 
alkali halide overlayers) could give rise to hysteretic phenomena, and in particular how they would
be connected to the presence of misfit superstructures ("solitons") . 
This is the issue which we address here by means of dynamical simulations
of the simplest one-dimensional model.
%
%
Our main result is the identification of an unexpected soliton-related hysteretic
mechanism.
During the first part of its swing, the tip can locally drag or push an
underlying defect -- here a soliton portion -- 
causing it to jump across a (Peierls-Nabarro) energy barrier. 
During the return journey, the defect 
follows only sluggishly, and remains trapped somewhat longer on the 
wrong side of the energy barrier, thus opening a hysteresis loop.
The area enclosed in the hysteretic force-displacement diagram represents a
large tip energy dissipation, one that can survive down to realistically
low AFM vibration frequencies, a regime where the linear-response
dissipation is quantitatively irrelevant.
This mechanism is likely to play a significant role every time a ``softness
pattern'' is present, and should be easier to observe for horizontal than
for vertical tip oscillations.
%

\section{The model}
%
To emphasize the basic and general aspects of the phenomena, rather than
a realistic model targeted on the chemical detail of a specific tip/sample
configuration \cite{Kantorovich04,Trevethan06,Ghasemi08,Caciuc08},
we use the simplest possible model -- a tip potential
oscillating over a one-dimensional harmonic chain (the overlayer) moving 
in a rigid incommensurate periodic potential (the substrate), Fig.~\ref{model:fig}c.
The Hamiltonian of the mobile overlayer atoms is
\begin{equation}\label{energytot}
H = E_k + U_{\rm at-at}+U_{\rm sub}+U_{\rm T}(t) \,,
\end{equation}
where $E_k = \frac m2 \sum_i \dot x_i^2$ is the kinetic energy,
\begin{equation}
U_{\rm at-at} = \frac{K}{2}\sum_{i}(x_{i+1}-x_{i}-a_{0})^{2} \;, 
\end{equation} 
is the mutual (harmonic) interaction potential, and 
$U_{\rm sub} = \sum_{i} v(x_{i})$ is the substrate potential, which we take of
a pure cosine form:  
\begin{equation}
v(x)=-\frac {F_{\rm sub}\,a_{\rm sub}}{4\pi} \cos(k_{\rm sub}\,x) \;.
\end{equation} 
%
%
%
Here $a_{0}$ is the mean spacing between adatoms, $K$ is their mutual
spring constant, $k_{\rm sub}= 2\pi/a_{\rm sub}$, and $a_{\rm sub}$ is the
period of the substrate potential \cite{Braunbook,VanossiJPCM}.
The two periodicities $a_{0}$ and $a_{\rm sub}$ define the coverage
$\theta=a_{\rm sub}/a_{0}$.
For convenience we take $a_{\rm sub}$ as the unit length, $F_{\rm sub}$ as
the force unit, and the mass $m$ of the particles as the mass unit.
%
(To get a feeling for quantities, 
the frequency units $F_{\rm sub}^{1/2}\, m^{-1/2}\, a_{\rm sub}^{-1/2}$ 
should be typical of an atomic vibration or a Debye frequency $\omega_{\rm D}$, 
typically 1~THz or more).
The general lack of commensuration between adsorbate and substrate
periodicities gives rise to two-dimensional misfit dislocations, sometimes
called solitons, which form a regular superstructure with the beat
periodicity between the two.
Fixed boundary conditions (BCs) are chosen 
in order to prevent the advancing tip to drag the entire pattern along,
that would occur if, e.g., periodic BCs were used instead.

$U_{\rm T}(t)=\sum_{i}u_{\rm T}(x_{i},t)$ is the time-dependent oscillating
potential describing the tip action on the overlayer.
We represent the AFM tip as a Gaussian-shaped oscillating potential,
with $u_{\rm T}(x,t)= u(x-x_{\rm T}(t)\,)$, $x_{\rm T}(t)=\bar x_{\rm T}+
\Delta_{\rm T}\cos(\omega_{\rm T}t)$, and
\begin{equation} \label{potenergytip}
u(x)= A_{\rm T}\,\exp\!\left(-x^2 / \sigma_{\rm T}^2\right) \;.
\end{equation}
Here $A_{\rm T}$ represents the repulsive (contact AFM, $A_{\rm T}>0$) or
attractive (noncontact AFM, $A_{\rm T}<0$) tip-atom interaction strength,
$\sigma_{\rm T}$ is the tip width, $\Delta_{\rm T}$ and $\omega_{\rm T}$
are the tip oscillation amplitude and angular frequency around its central
position $\bar x_{\rm T}$.

The equation of motion for the $i$-th overlayer atom is
\begin{equation} \label{motionequation}
m \ddot x_{i} = -v'(x_{i}) + K (x_{i+1}+x_{i-1}-2x_{i}) + f_{\rm T} (x_{i},t) - \gamma \dot x_{i}
\,,
\end{equation}
where $v'(x)=\frac 12 F_{\rm sub}\,\sin(k_{\rm sub}x)$,
and the tip force
\begin{equation} \label{tipforce}
f_{\rm T}(x,t)= - \frac \partial{\partial x} u_{\rm T}(x,t) 
\end{equation}
is given by a straightforward analytical expression.
A damping force term $-\gamma \dot x_i$ is introduced to represent all
dissipation phenomena which remove energy 
and allow the attainment of a stationary frictional state.

We integrate the equations of motion (\ref{motionequation}) by means of a
standard adaptive fourth-order Runge-Kutta routine \cite{NumericalRecipes}
starting each simulation from a stationary fully relaxed overlayer, as
obtained by a preliminary relaxation of equally-spaced adatoms
$x_i(0)=i\cdot a_0$ and $\dot x_i(0)=0$.
All simulations are carried out at a nearly commensurate coverage
$\theta=1.06=\frac{53}{50}$, realized by means of a chain of $N=107$
particles in a region of length $L=100\, a_{\rm sub}$.
A finite temperature $T$ is implemented
by adding a standard Langevin random force
to Eq.~(\ref{motionequation}), and averaging over a long simulation time,
usually at least 100 tip-oscillation periods.
The extreme simplicity of the model allows us to extend simulations down to the
realistic AFM frequency in the MHz range, which requires exceedingly 
long integration times.

The instantaneous power drained away by the damping term amounts to
\begin{equation}\label{power}
P_{\rm diss}=\sum_i  \dot x_i \cdot (\gamma \dot x_i)
=\gamma \sum_i \dot x_i^2=\frac{2 \gamma}{m}\,E_c \,,
\end{equation}
and is thus proportional to the total kinetic energy of the overlayer.
The power pumped by the tip into the chain is
\begin{equation}\label{powertip}
P_{\rm T}=\sum_i  f_{\rm T}(x_i,t)\cdot\dot x_i \,.
\end{equation}
While these two quantities fluctuate separately, they must of course
coincide on average over a period $\tau=2\pi/\omega_{\rm T}$ in the dynamical
steady state
\begin{equation} \label{powertipaverage}
\overline P = \frac 1\tau \int_\tau \! dt\, P_{\rm T}(t)
= \frac 1\tau \int_\tau\! dt \, P_{\rm diss}(t) \,,
\end{equation}
also indicating how the work done by the tip oscillation is eventually
dissipated entirely by the viscous friction term.

\section{Results: linear response and beyond}
%
\begin{figure}
\centerline{
\includegraphics[width=82mm,angle=0,clip=]{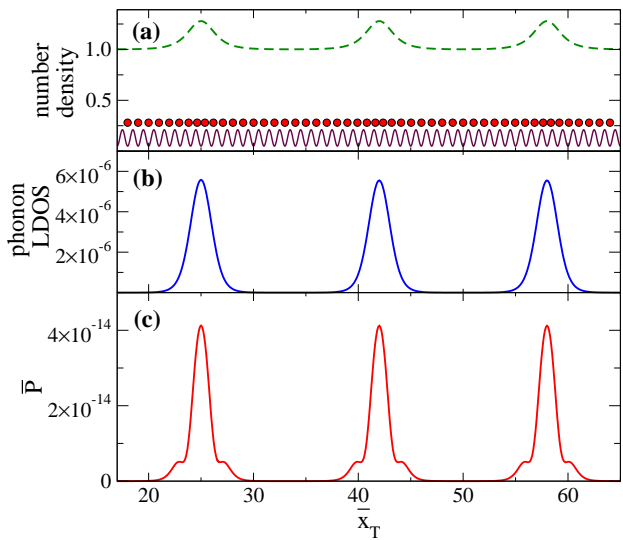}
}
\caption{\label{linear_regime:fig} (Color online)
The linear-response regime.
(a) Overlayer atom density (dashed), rest positions (circles), in
the periodic substrate potential (solid).
(b) The density of zero-frequency overlayer vibrational modes peaks
at the soliton positions.
(c) Dissipated power during tip vibration for a tip width 
$\sigma_{\rm T}=\Delta_{\rm T}=a_{\rm sub}$, weakly attractive potential of amplitude
$A_{\rm T}=-10^{-4}$, $\gamma=0.2$ (underdamped regime), $K=5$ (fairly
rigid overlayer), and oscillation frequency $\omega_{\rm T}=10^{-4}\, \pi$
(roughly 1~GHz). This regime is described well by linear-response theory.
Note a huge dissipation peak contrast at the soliton position relative to
the terrace between solitons of about $10^{4}$.
}
\end{figure}

Figure~\ref{linear_regime:fig} displays the dissipation results obtained in
simulation for a weakly attractive tip potential, very high oscillation
frequency, and general parameters that fall well inside the linear-response
regime \cite{Gauthier99,Gauthier00, Granato99,PerssonBook,PerssonTosatti99}.
The linear response results show (i) strong dissipation enhancement at
solitons, with $\bar P(\bar x)$
several orders of magnitude stronger than in a terrace between two of them, closely mirroring the
phonon local density of states (LDOS);
(ii) dissipated power which is proportional to $A_{\rm T}^2$, independent of the
attractive/repulsive sign of the tip-overlayer interaction, i.e.\ of 
the noncontact or contact mode of the AFM; (iii) absolute dissipation 
values that are very weak everywhere, 
and dropping with decreasing AFM frequency as $(\omega_{\rm T}/\omega_{\rm D})^2$.
Summing up, the predicted relative contrast of the soliton pattern in
linear response dissipation is indeed very large.
However, the exceedingly low value of realistic AFM frequencies
($\lesssim\,$MHz) relative to microscopic frequencies ($\sim\,$THz) renders
this linear-dissipation mechanism entirely academic.


\begin{figure*}
\centerline{
\hfill
{\small a}
\includegraphics[height=61mm,angle=0,clip=]{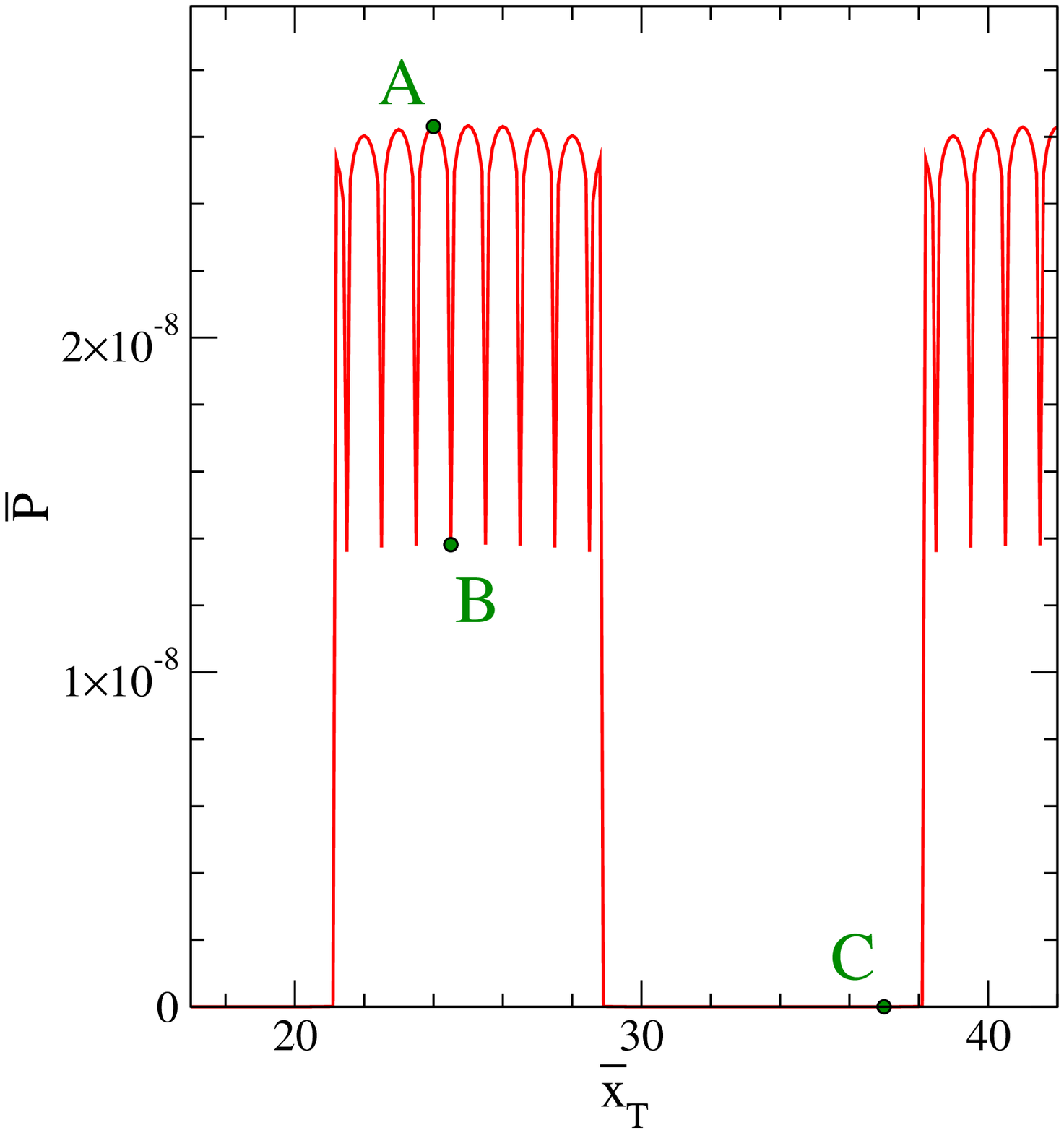}
\hfill
{\small b} 
\includegraphics[height=61mm,angle=0,clip=]{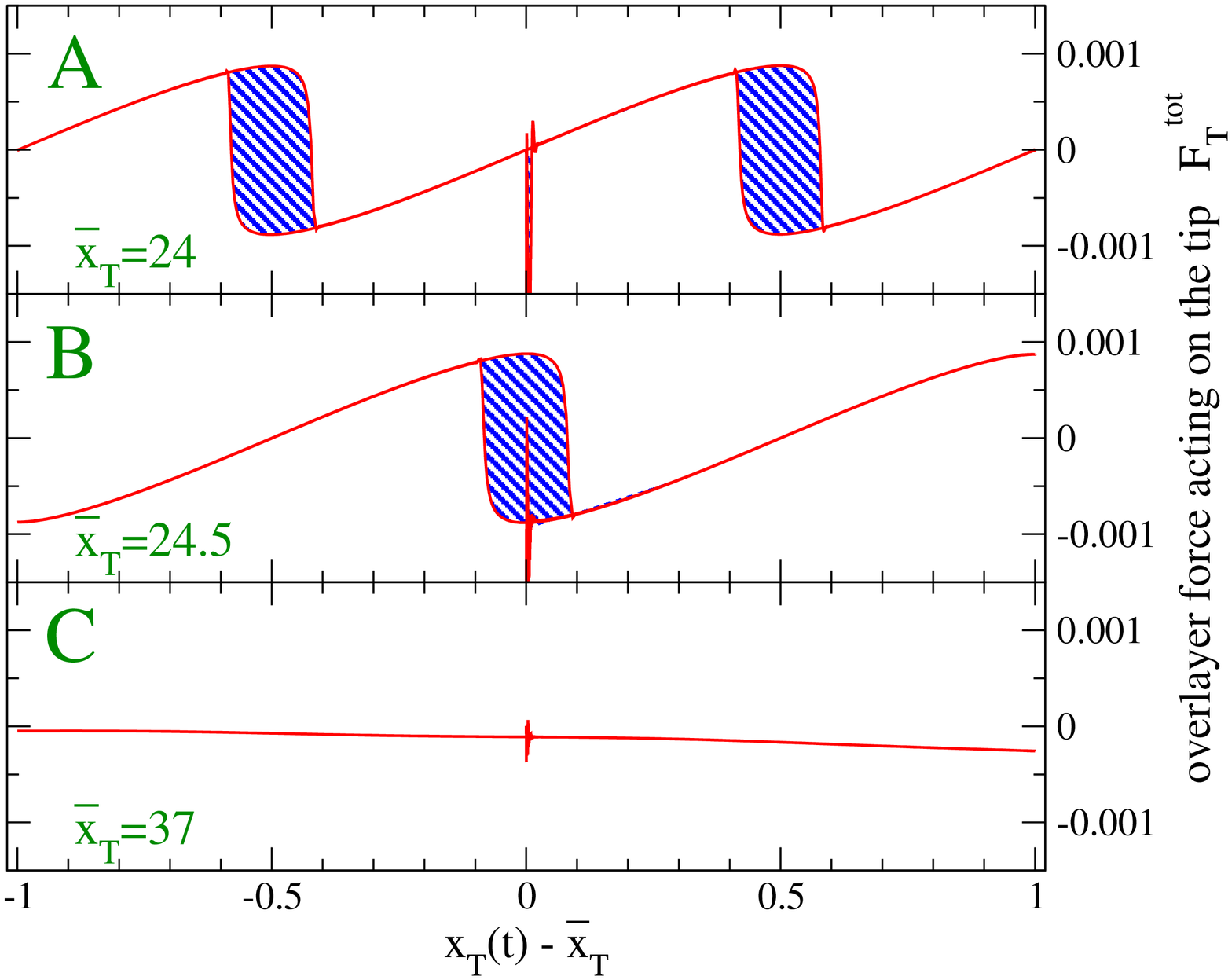}
\hfill
}
\caption{\label{nonlinear:fig} (Color online)
The strong-interaction hysteretic regime.
(a) Mean power dissipated in the steady regime by a 
strongly interacting tip ($A_{\rm T}=-0.01$, all other parameters the same
as in Fig.~\ref{linear_regime:fig}) scanning the same overlayer ($K=5$).
(b) Force-displacement response at three typical scanning points
marked in the left panel. The strong dissipation at the
solitons is now due to hysteretic jumps of the solitons (kinks) across
their Peierls-Nabarro barriers.
The dissipated energy in a cycle equals the area of the
hysteretic loop in the force-displacement plane (shaded).
When the potential is strong enough to drag or push a soliton (points A, B),
this occurs with hysteresis and the dissipation is large.
With the selected oscillation amplitude, the soliton
is dragged across two barriers (point A) or a single barrier (point B),
depending on the center of tip position $\bar x_{\rm T}$.
When the tip grabs no soliton (point C) there is no hysteresis and
dissipation drops.
}
\end{figure*}

\begin{figure}
\centerline{
\includegraphics[width=82mm,angle=0,clip=]{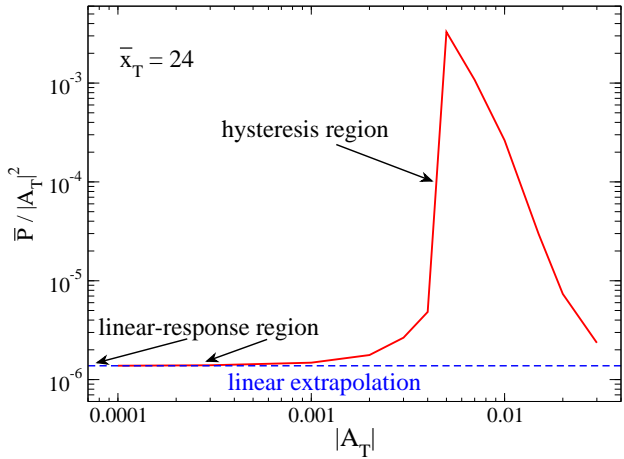}
}
\caption{\label{nonlinearity:fig} (Color online)
The dissipated power normalized to the linear regime scaling,
$\overline{P}/|A_{\rm T}|^2$, versus the 
strength of the tip-overlayer interaction
$|A_{\rm T}|$, at point A of Fig.~\ref{nonlinear:fig} and for the same
parameters as in Fig.~\ref{linear_regime:fig}.
Notice the sharp onset of the non-linear regime where the hysteretic
mechanisms starts to play a role, leading to a strong enhancement of
dissipation.
}
\end{figure}

We reach a more realistic regime by enhancing the tip-overlayer interaction
strength, while still remaining in a moderate-interaction regime
representing noncontact AFM.
This new regime is dominated by nonlinear effects, where dissipation no
longer drops as $\omega_{\rm T}^2$, but at most linearly in $\omega_{\rm
  T}$ (apart from logarithmic corrections).
Comparison of Fig.~\ref{nonlinear:fig}a with Fig.~\ref{linear_regime:fig}c
shows that, nonlinear dissipation is again much larger near the solitons
than in between them.
A two-order of magnitude increase in $A_{\rm T}$ would in linear regime
imply a dissipation increase by a factor $10^4$, whereas we find a much
larger factor of about $10^6$ already at this large frequency 
($\omega_{\rm T}$ is here $10^{-4}\pi$, corresponding to the gigahertz range).
%
Decreasing frequency down to realistic AFM values, the increase will become
gigantic, because the nonlinear dissipation lacks the extra power of
$\omega_{\rm T}$ appearing in the linear-regime dissipation.
The new element brought in by nonlinearity is mechanical. 
A strongly interacting tip is now able to drag, or to push, the soliton
-- a mobile entity -- forward or backward during the oscillation cycle.
As the soliton must overcome the (Peierls-Nabarro) barrier in order to
move, its motion is sluggish, and can follow the tip only with hysteresis
and, as anticipated, hysteresis entails a large dissipation.
As shown in detail in Fig.~\ref{nonlinear:fig}a,b, the higher dissipation
point A is found to corresponds to two successive Peierls-Nabarro barriers
being overcome in the oscillatory process, the smaller dissipation point B
to a single barrier.
The dissipation at point C, where the tip potential is unable to ``grab''
the soliton, is negligible by comparison.
The onset of this large-dissipation region, dominated by hysteresis, is
rather sharp.  Fig.~\ref{nonlinearity:fig} illustrates this point, by
showing the average power $\overline P$ at location A, divided by the
linear-regime factor $|A_{\rm T}|^2$, as a function of the tip amplitude
$|A_{\rm T}|$.
Beyond a value of $|A_{\rm T}|$ of order $4\times 10^{-3}$, the linear
regime behavior is abruptly abandoned, and the dissipation increases
rapidly by several orders of magnitude.

\begin{figure}
\centerline{
\includegraphics[width=82mm,angle=0,clip=]{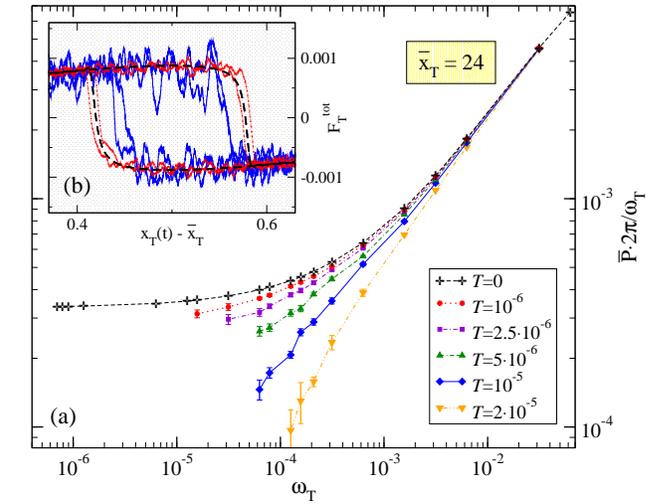}
}
\caption{\label{powerVSfrequency:fig} (Color online)
The dissipation reduction due to thermal shrinking of the hysteretic loop,
for $\omega_{\rm T} =10^{-4}\,\pi$ at the soliton location A of
Fig.~\ref{nonlinear:fig}.
(a) The frequency dependence of the energy dissipated in one period 
$\bar P \, 2\pi/\omega_{\rm T}$, computed for several temperatures,
exhibiting a
clear reduction of $\bar P \, 2\pi/\omega_{\rm T}$ due to a rise in
temperature.
(b) Detail of the right-side area of the force-displacement dependency
following two typical tip-oscillation periods for $T=10^{-5}$, solid, and
$10^{-6}$, dotted, compared to $T=0$, dashed: the hysteretic area is
shrinking due to a randomly anticipated thermally activated barrier
crossing.
}
\end{figure}

\section{Predictions and discussion}
%
Our simulated example strongly suggests that a large hysteretic component
should be present in the existing dissipation maps \cite{Albers09} of
Moir\'e patterns.
More generally, hysteretic defect dragging should dominate the AFM
dissipation maps.
What are the predicted signatures of this mechanism?
Our model study suggests two main signatures.

(i) Abruptness of AFM friction onset with increasing strength of
tip-surface interaction.
As suggested by Figs.~\ref{nonlinear:fig}-\ref{nonlinearity:fig}, dragging
sets in abruptly only above a certain threshold, which means below a
certain tip-surface or tip-soliton distance.

(ii) Anomalously mild (linear with logarithmic corrections) frequency
dependence of AFM friction at finite temperature.
It is a rather general property of all hysteretic friction phenomena to
heal away at sufficiently low frequencies, where adiabatic motion allows
sufficient time to jump thermally over barriers.
For instance, thermolubricity experiments
\cite{Gnecco00,Riedo03,Gnecco03,Sills03} and detailed calculations within
the Tomlison model \cite{Sang01} show an average friction force
$F=F_c - A \,T^{2/3} \left|\log(\alpha v/T)\right|^{2/3}$, where $v$ is the
driving velocity, and $A$ and $\alpha$ are system-dependent dimensional
constants.
In incommensurate overlayers, the soliton nearest to the tip behaves
similarly to a Tomlinson particle, as is driven across a Peierls-Nabarro
barrier.
In Fig.~\ref{powerVSfrequency:fig}a we do observe a thermal
reduction of dissipation, due to a shrinkage of the hysteretic loop,
illustrated in Fig.~\ref{powerVSfrequency:fig}b.
When driving is oscillatory as in AFM dissipation, the role of $v$ is taken
by $\Delta_{\rm T}\,\omega_{\rm T}$.
We find our data to be compatible with a similar relation 
$\overline{P}(T)=\overline{P}(0)
-A'\, T^{2/3} \left|\log( \alpha'\omega_{\rm T}/T) \right|^{2/3}$.
The parameters $A'$ and $\alpha'$ are here related to the effective soliton
properties (mass, damping, barrier height...), and are nontrivial functions
of the ``bare'' model parameters.


We conclude that AFM dissipation maps of incommensurate overlayer
superstructures can in principle achieve an extremely high contrast
resolution of soliton defects relative to commensurate terraces.
The most important theoretically predicted dissipation mechanism is the nonlinear
dragging or pushing of some local portion of the defect, 
where the large tip damping is associated with hysteresis of defect motion.
Besides a sharp threshold in the tip-surface interaction and oscillation
magnitudes, this mechanism predicts a very characteristic logarithmic
dependence (eventually turning to linear at extremely low frequencies) of dissipation upon frequency and temperature.
More generally, the nonlinear dragging of soft defects or features (e.g.\ a
floppy residue in a biomolecule) should give rise to a strong visibility in
AFM dissipation topography, of considerable potential impact for applications.

\acknowledgments
This work was supported by CNR, as a part of the
European Science Foundation EUROCORES Programme FANAS.



\end{document}